\documentstyle [12pt,epsf]{article}
\begin{document}
\newcommand{\vm}{\vspace{0.2cm}}
\newcommand{\vl}{\vspace{0.4cm}}
%123456789%123456789%123456789%123456789%123456789%123456789%123456789%123456789

\title{p-Adic TGD: Mathematical ideas.}

\author{Matti Pitk\"anen
\\
\small \em, Torkkelinkatu 21 B 39,  00530 Helsinki, Finland\\
}

\date{30. June 1995}

\maketitle

\newpage

\tableofcontents

\newpage

\begin{center}
Abstract
\end{center}

\vl

The mathematical basis of p-adic Higgs mechanism discussed in papers
hep-th@xxx.lanl.gov 9410058-62  is considered in this paper. The basic
properties of p-adic numbers, of  their algebraic extensions and the so
called canonical identification between positive real numbers and p-adic
numbers are described.  Canonical identification induces p-adic topology and
differentiable structure on real axis  and  allows definition of definite
integral with physically desired properties.  p-Adic numbers together with
canonical identification provide analytic tool to produce fractals.
Canonical identification makes it possible to generalize probability
concept, Hilbert space concept,  Riemannian metric and Lie groups to p-adic
context.  Conformal invariance generalizes to arbitrary dimensions since
p-adic numbers allow algebraic extensions of arbitrary dimension. The
central theme of all developments is the existence of square root, which
forces unique quadratic extension having dimension $D=4$ and $D=8$ for $p>2$
and $p=2$ respectively. This  in turn implies that the dimensions of p-adic
Riemann spaces are multiples of $4$ in $p>2$ case and of $8$ in  $p=2$ case.

\newpage

Note:

The .eps files representing p-adic fractals discussed in the text as well as
MATLAB programs needed to generate the fractals are supplied by request.
The commands attaching .eps files to text are in the text but preceided
by comment signs: please remove these signs.

\newpage
\section{Introduction}

 There are a lot of  speculations about the role of p-adic
 numbers in Physics \cite{Padstring,Narstrings,Padvira}. In \cite{padrev}
 one can find
 a review
of the work done.     In general the work is related to quantum theory and
based on
assumption
 that quantum mechanical state space is ordinary complex Hilbert space.
 This is not
 absolutely
 necessary since p-adic unitarity and probability concepts make sense
 \cite{padprob}.
 What
is however essential is some kind of correspondence between p-adic and real
numbers
since the
predictions of, say, p-adic quantum mechanics should be expressed in terms
of real
numbers.
The formulation of physical theory using p-adic state space and p-adic
dynamical
variables
requires also the construction of p-adic differential and integral calculus.
Also
the p-adic
counterpart of Riemann geometry as well as group theory  are needed.  In
this  chapter
the aim is
to carry out these generalizations.

\vm

 The key
observation behind all developments to be represented in the sequel is very
simple: there
is
canonical correspondence between p-adic numbers and nonnegative real
numbers given
by "pinary"
expansion of real number: positive real number $x= \sum x_np^n$
($x=0,1,..,p-1$, $p$
prime) is
mapped to  p-adic number $\sum  x_np^{-n}$. This canonical correspondence
allows to
 induce
p-adic topology and differentiable structure to the real axis.
p-Adically
differentiable
functions define  typically  fractal like real  functions via canonical
identification so
that p-adic numbers provide analytic tool for producing fractals.
p-adic numbers allow
algebraic
extensions of arbitrary dimension and the concept of complex analyticity
generalizes to
p-adic
analyticity.  The fact that  real continuity implies p-adic continuity
implies that real
physics can emerge  above some  length scale  $L_p$ as an excellent
approximation
of underlying p-adic physics.

 \vm

The  canonical correspondence makes possible to  generalize the
concepts of inner product, integration,  Hilbert space,  Riemannian metric,
Lie group
theory and
Quantum mechanics to p-adic context in a relatively straightforward
manner.
Essentially the
fractal counterparts of all these structures  are obtained in this
manner.
A possible
reason for the practical absence of p-adic physics  is  probably  that the
existence and
importance of the  canonical correspondence has not been realized.  The
successfull p-adic
description of Higgs mechanism relies heavily on canonical correspondence.
In later
chapters it will be found that the concepts of p-adic probability and
unitarity make sense and
one can associate with  p-adic probabilities unique real probabilities
using canonical
correspondence and this
 predicts novel physical effects.

\vm

The topics of  the  chapter are following:  \\
i) p-Adic numbers,  their algebraic extensions and  canonical identification
are
described. The existence of square root of p-adically real number
is necessary
in many
applications of p-adic numbers (p-adic group theory,  p-adic unitarity,
Riemannian
geometry) and
its existence implies  unique algebraic extension, which is
4-dimensional in  $p>2$
case and
8-dimensional in $p=2$ case.\\ ii)   p-Adic valued  inner product
necessary for
various
generalizations   is introduced. \\ iii) The concepts of p-adic
differentiability and
analyticity are introduced and
 the fractal properties of p-adically
 differentiable functions as well as nondeterminism of
 p-adic differential equations  are demonstrated. It is also shown  that
 period
doubling property is characteristic feature of 2-adically
 differentiable functions. \\
 iii)  The concept of  p-adic valued integration is defined: this concept
 is necessary
in order to formulate p-adic variation principles.  \\
iv)  formulate p-adic Riemannian  needed in TGD: eish applications:  the
existence of p-adic inner product and p-adic valued integration is essential
for these
developments.  The dimensions of p-adic Riemann spaces are multiples
of $4$ ($p>2$)
or $8$
($p=2$).   It is hardly an accident that these dimensions are spacetime
and imbedding
space
dimensions in TGD. \\ v)  consider some characteristic details related to
p-adic counterparts of
Lie-groups

\section{p-Adic numbers}

p-Adic numbers ($p$ is prime: 2,3,5,... ) can
be regarded as a
completion
of rational numbers using norm which, is different from
 the ordinary
norm of real numbers \cite{2adic}.   p-Adic numbers are
 representable
as power
expansion of the prime number $p$ of form:\\

\begin{eqnarray}
x&=& \sum_{k \ge k_0} x(k) p^k ,  \ x(k) =0,....,p-1
\end{eqnarray}

\noindent
The norm of a p-adic number given by

\begin{eqnarray}
\vert x \vert &= &p^{-k_0(x)}
\end{eqnarray}

\noindent Here $k_0(x)$ is the lowest power in the expansion of p-adic
number.
The norm differs drastically from the norm of ordinary real numbers since
it depends on
the
lowest pinary digit of the p-adic number only. Arbitrarily high powers in
the expansion
are
possible since the norm of p-adic number is finite also for numbers, which
are infinite
with
respect to the ordinary norm. A convenient  representation for  p-adic
numbers is in
 the
form

\begin{eqnarray}
x&=& p^{k_0} \varepsilon (x)
\end{eqnarray}

\noindent  where $\varepsilon (x)=k+....$
with $0<k<p$,  is
p-adic number with unit
norm and
analogous
to
the phase factor $exp(i\phi)$ of complex number.

\vm

The distance function $d(x,y)= \vert x-y\vert_p$ defined by p-adic norm
possesses a very general property called  ultrametricity:

\begin{eqnarray}
d(x,z)&\le&max\{ d(x,y),d(y,z)\}
\end{eqnarray}

\noindent The properties of the distance function make it possible to
decompose
the  $R_p$
into a union of disjoint sets using the criterion  that $x$ and $y$
belong to same
class if the
distance between $x$ and $y$ satisfies the condition

\begin{eqnarray}
d(x,y)&\le& D
\end{eqnarray}

\noindent  This division of the metric space into classes has following
properties:\\
a) Distances between the members of two different classes $X$ and $Y$ do
not depend on
the choice of points $x$ and $y$ inside classes. One can therefore speak
about distance
function
between classes. \\ b) Distances of points $x$ and $y$ inside single
class are smaller than
distances between different classes.  \\ c)  Classes form a hierarchical
tree.

\vm

Notice that the  concept of ultrametricity emerged to Physics in
the models for spin
glassess and is believed to have also applications in biology \cite{Parisi}.
The
emergence of
p-adic topology as effective topology of spacetime would make ultrametricy
property basic feature
of Physics at long length scales.

\section{Canonical correspondence between p-adic and real numbers}

There exists a
natural continuous
map $Id: R_p \rightarrow R_+$  from p-adic
 numbers to
non-negative  real numbers
given by the "pinary" expansion of the real number for
 $x\in R$ and $y \in R_p$ this correspondence reads

\begin{eqnarray}
y&=& \sum_{k>N} y_k p^k\rightarrow x=\sum_{k<N} y_k p^{-k}
\nonumber\\
y_k &\in& \{0,1,..,p-1\}
\end{eqnarray}

\noindent  This map is continuous as one easily finds out. There is however
a
little difficulty associated with the definition of the inverse map since
the
pinary expansion
like also desimal expansion is not unique ($1= 0.999...$) for real numbers
$x$,
which allow
pinary expansion with  finite number of pinary digits

\begin{eqnarray}
x&=&\sum_{k=N_0}^{N} x_k p^{-k}\nonumber\\
x&=& \sum_{k=N_0} ^{N-1} x_k p^{-k}+
(x_N-1) p^{-N} +(p-1)p^{-N-1}\sum_{k=0,..} p^{-k} \nonumber\\
\
\end{eqnarray}

\noindent The p-adic images associated with these expansions are different

\begin{eqnarray}
y_1&=&\sum_{k=N_0}^{N} x_k p^{k}\nonumber\\
y_2&=& \sum_{k=N_0}^{N-1} x_k p^k+
(x_N-1) p^{N} +(p-1)p^{N+1}\sum_{k=0,..} p^{k} \nonumber\\
&=& y_1 + ( x_N-1)p^N -p^{N+1}
\end{eqnarray}

\noindent so that the inverse map is either two-valued for p-adic numbers
having
expansion with finite pinary digits or single valued and discontinuous
and
nonsurjective
if one makes pinary expansion unique by choosing the one with finite pinary
digits.
The finite
pinary digit expansion is natural choice since in applications one
always must use pinary cutoff in real axis. Furthermore. p-adicity is good
approximation only below (rather than above, as thought originally) some
length scale, which means pinary cutoff on real axis.

\vm

What about the  p-adic counterpart of negative real numbers?
In  TGD:eish applications this correspondence is not needed since canonical
identification is
used only in the direction $R_p \rightarrow R$. Furthermore, it is always
possible to choose
the real coordinates of finite spacetime region so that coordinate variables
are nonnegative
so that the problem disappears.

 \vm

The canonical  identification map will be crucial for the proposed
applications of
p-adic numbers. The  topology induced by this map  in the set of
positive real
numbers
differs from ordinary topology.  The difference is easily understood by
interpreting
the p-adic
norm as a norm in the set of real numbers. The norm is constant in each
interval
$[p^k,p^{k+1})$  (see Fig. \ref{Norm}) and is equal to the usual real
norm at the
points $x=
p^k$: the usual linear norm is replaced with a piecewise constant norm.
This means
that
  p-adic topology is coarser than the usual real topology and
the higher the value of $p$ is,  the coarser the resulting topology is
above given length
scale. This  hierarchical ordering of p-adic topologies will be
central feature
as far as
the proposed  applications of the p-adic numbers are considered.

\vm

Ordinary continuity implies p-adic continuity since the norm induced from
p-adic
topology is
rougher than  ordinary norm. This means that physical system can be
genuinely p-adic below
certain length scale $L_p$ and become in good approximation real, when
length scale resolution
$L_p$ is used in its description. TGD:eish applications rely  on this
assumption. p-Adic
continuity implies  ordinary continuity from right   as is clear
already from the
properties of p-adic norm (the graph of the norm is  indeed continuous
from right).  This
feature is one clear signature of p-adic topology.

%\begin{figure}
%\leavevmode
%\centering
%\vspace*{1cm}
%\epsfxsize=5 cm \epsfysize=5 cm \epsfbox{norm.eps}
%\label{Norm}
%\caption{The real norm induced by canonical identification from 2-adic norm.}
%\end{figure}

The   linear  structure of p-adic numbers induces corresponding
structure in the set
of positive real numbers and p-adic linearity in general differs from
the ordinary
concept of
linearity. For example, p-adic sum is equal to real sum only provided
the summands
have no
common pinary digits. Furthermore, the condition $x+_p y <max\{x,y\}$
holds in general
for the
p-adic sum of real numbers.  p-Adic multiplication is equivalent with
ordinary
multiplication
only provided  that  either of the members of the product is power of $p$.
Moreover
one has $x
\times_p y <x \times y$ in general. The p-Adic negative $-1_p$ associated
with p-adic
unit 1 is
given by  $(-1)_p= \sum_k (p-1) p^k$  and defines p-adic negative for each
real number $x$.  An interesting
possibility is that
p-adic linearity might replace ordinary linearity in strongly nonlinear
systems so that
nonlinear systems would look simple in p-adic topology.

 \vm

Canonical correspondence is quite essential in TGD:eish applications.
Canonical identification makes it  possible  to define p-adic valued
definite integral  and this is cornerstone of  TGD:eish applications.
Canonical identification is in key role in the   successfull predictions
of the elementary particle masses.   Canonical identification  make also
possible  to undestand the  connection between p-adic and real probabilities.
These
and many
other succesfull applications  suggests that canonical identification
is involved with some deeper mathematical structure.  The following
inequalities hold true:

\begin{eqnarray}
(x+y)_R&\leq& x_R+y_R\nonumber\\
\vert x\vert_p \leq (xy)_R&\leq &x_Ry_R
\end{eqnarray}

\noindent where  $\vert x \vert_p$ denotes p-adic norm.  These inequalities
 can be generalized to case of $(R_p)^n $ ( linear space over p-adic
 numbers).

\begin{eqnarray}
(x+y)_R&\leq& x_R+y_R\nonumber\\
\vert \lambda \vert_p\vert y\vert_R \leq (\lambda y)_R&\leq &\lambda_Ry_R
\end{eqnarray}

\noindent where the norm of the vector $x\in T_p^n$ is defined in some
manner.
 The case of Euclidian space suggests the definition

\begin{eqnarray}
(x_R)^2 &=& (\sum_n x_n^2 )_R
\end{eqnarray}

\noindent   These inequilities resemble those satisfied by vector norm.
The only difference is failure of linearity in the sense that the norm of
scaled
vector is not obtained by scaling the norm of original vector.  Ordinary
situation prevails  only   if scaling corresponds to power  of $p$.
Amusingly,
the p-adic counterpart of Minkowskian norm

\begin{eqnarray}
(x_R)^2 &=& (\sum_k x_k^2- \sum_l x_l^2 )_R
\end{eqnarray}

 \noindent  produces nonnegative  norm.
Clearly the p-adic space with this norm  is analogous to future light cone.

\vm

These observations suggests that the concept of normed space or Banach
space
might have generalization and physically the generalization might apply
to the
description of nonlinear system. The nonlinearity would be concentrated
in the
nonlinear behaviour of the norm under scaling.

\section{Algebraic extensions of p-adic numbers}

Real numbers allow only complex numbers as an
 algebraic extension.
 For p-adic numbers algebraic extensions of
 arbitrary dimension are possible \cite{2adic}.  The simplest manner to
 construct
 (n+1)-dimensional extensions is to consider irreducible polynomials
 $P_n(t)$ in $R_p$
 assumed to
have rational coefficients:  irreduciblity means that polynomial does not
possess roots
in $R_p$
so that one cannot decompose it into a product of lower order $R_p$ valued
polynomials.
Denoting
one of the  roots of $P_n(t)$ by $\theta $ and defining $\theta^0= 1$  the
general
form of the
extension is given by

\begin{eqnarray}
Z&=& \sum_{k=0,..,n-1}x_k \theta^k
\end{eqnarray}

\noindent  Since $\theta$ is root of the polynomial in $R_p$ it follows
that $\theta^n$ is
expressible as sum of lower powers of $\theta$ so that  these numbers
indeed form n-dimensional
linear space with respect to the p-adic topology.

\vm

 Especially simple  odd dimensional extensions are
  cyclic extensions  obtained by
 considering the roots of the polynomial

\begin{eqnarray}
P_n(t) &=&t^n+ \epsilon d\nonumber\\
\epsilon&=& \pm 1
\end{eqnarray}

\noindent    For   $n=2m+1$ and $(n=2m, \epsilon=+1)$    the irreducibility
of $P_n(t)$ is
guaranteed   if $d$ does not possess $n$:th root in $R_p$.  For
$(n=2m,\epsilon=-1)$
one must
assume that $d^{1/2}$ does not exist p-adically.      In this case
$\theta$ is one of
the roots
of the equation

\begin{eqnarray}
t^n &=& \pm d
\end{eqnarray}

\noindent where $d$ is p-adic integer with finite number of pinary digits.
It is possible although not necessary to identify roots as complex numbers.
There
exists
$n$ complex roots of $d$
and $\theta$ can be chosen to be  one of the real or complex roots
satisfying the
condition
$\theta^n =\pm d$.  The roots can be written in the general form

\begin{eqnarray}
\theta& = &d^{1/n} exp(i\phi (m)), \ m=0,1,....,n-1\nonumber\\
\phi (m)&= &\frac{m2\pi}{n} \ or \ \frac{m\pi}{n}
\end{eqnarray}

\noindent Here $d^{1/n}$ denotes the real root of the equation
$\theta^n=d$.
Each of the phase factors $\phi (m)$ gives rise to algebraically equivalent
extension:
only the representation is different. Physically these extensions need not
be
equivalent since
the identification of p-adic numbers with complex numbers plays fundamental
role
in the
applications.  The cases $\theta^n = \pm d$ are physically and
mathematically quite
different.

\vm

The  norm of an algebraically extended p-adic number $x$ can be
defined
as
some power of the ordinary   p-adic norm of the determinant of the
linear map $ x: ^eR_p^n \rightarrow ^eR_p^n$ defined by the
multiplication with $x$: $y \rightarrow xy$

\begin{eqnarray}
N(x) &=& \vert det(x)\vert^{\alpha}, \ \alpha >0 \nonumber\\
\
\end{eqnarray}

\noindent  The requirement that norm is
 homogenous function of degree one in the components of the
algebraically
extended 2-adic number (like also the standard norm of  $R^n$ )
implies the condition
$\alpha=1/n$, where $n$ is the dimension of the algebraic extension.

\vm

The canonical correspondence between the points of $R_+$ and  $R_p$
generalizes
in obvious manner:  the point $\sum_k x_k\theta ^k$ of algebraic extension
is identified as the
point  $ (x^0_R, x^1_R,...,x^k_R, ..,)$ of $R^n $ using the
pinary
expansions of
the components of p-adic number.   The p-adic linear structure of the
algebraic
extension induces
linear structure in $R_+^n$ and p-adic multiplication induces multiplication
for
the vectors of
$R_+^n$. An exciting possibility is that p-adic linearity might replace
ordinary
linearity in
strongly nonlinear systems.

\section{Algebraic extension allowing square root on p-adic real axis}

The existence of square root of «real« p-adic  number is a common theme in
various
applications of p-adic numbers.  \\
a) The   p-adic generalization of the representation theory of ordinary
groups and
Super Kac Moody and Super Virasoro algebras exists provided an extension
of p-adic
numbers allowing square roots  of «real« p-adic numbes is used. The reason
is that matrix
elements of the raising and lowering operators in Lie-algebras as well as
oscillator
operators typically involve square roots.   \\ b) The existence of square
root of «real«
p-adic number is also necessary ingredient in the definition of
p-adic unitarity and
quantum probability  concepts  since the solution of the requirement
that  $p_{mn}=
S_{mn}\bar {S}_{mn} $ is p-adically real leads to expressions involving
square roots. \\
c)   p-Adic Riemannian geometry   necessitates the existence of
square root of
«real« p-adic numbers  since the definition of the infinitesimal length
$ds=
\sqrt{g_{ij}dx^idx^j} $  involves square root. \\ What is important is
that only the
square root of p-adically real numbers is needed:  the square root need
not exist outside
the real axis.   It is indeed impossible to find finite dimensional
extension allowing
square root for all numbers of the extension.
 For  $p>2$ the minimal dimension for algebraic extension allowing square
 roots near
 real axis is $D=4$.  For $p=2 $
the dimension of the extension is $D=8$.

\vm

For $p>2$ the form of the  extension can be derived by the following
arguments. \\
a)  For $p>2 $ p-adic number $y$  in the range $(0,p-1)$ allows  square
root only provided
there exists p-adic number $x\in \{0,p-1\}$ satisfying the condition
$y= x^2 \  mod \ p$.
Let
$x_0$ be the smallest integer, which does not possess p-adic square root
and add the square
root
$\theta$ of $x_0$ to the  number field.   The numbers in the extension
are of the form $x+
\theta y$. The extension allows square root for every $x\in \{0,p-1\}$ as
is easy to see.
 p-adic numbers $mod \ p$ form a finite field $G(p,1)$ \cite{2adic} so that any
 p-adic number $y$,
 which does not possess square root can be written in the form $ y= x_0 u$,
 where $u$
possesses square root. Since $\theta$ is by definition the square root
of $x_0$ then also
$y$ possesses square root.   The extension does not depend on the
choice of $x_0$.

\vm

The square root of $-1$ does not exist for $ p \  mod \ 4 = 3$
\cite{Number} and $p=2 $
but the
addition of $\theta$ gurantees its existence automatically.  The
existence of $\sqrt{-1}$
follows
from the existence of $\sqrt{p-1}$ implied by the extension by   $\theta$.
$\sqrt{(-1+p) -p}$
can  be developed in power in powers of $p$ and series converges since the
p-adic norm of
coefficients in Taylor  series is not larger than $1$.  If $p-1$ doesn not
possess square
one can
take $\theta$ to be equal to $\sqrt{-1}$.

\vm

b) The next step is to add square root of $p$ so that extension becomes
4-dimensional and
arbitrary number in the extension can be written as

\begin{eqnarray}
Z&=& (x+\theta y) +\sqrt {p}( u +\theta v)
\end{eqnarray}

\noindent  This extension is natural for p-adication of spacetime surface
so that
spacetime
can be regarded as a number field locally.
An important point to notice that the extension guarantees the existence
of square for
«real« p-adic numbers only.

\vm

c) In $p=2$ case 8-dimensional extension is needed to define square roots.
  The addition of $\sqrt{2}$ implies that one can restrict the
consideration
 to the square roots of odd 2-adic numbers.  One must be careful in
defining
 square roots by the Taylor expansion of square root $\sqrt{x_0+x_1} $
 since
  $n$:th Taylor coefficient is  proportional to $2^{-n}$  and possesses
 2-adic norm $2^n$.   If $x_0$ possesses norm $1$ then $x_1$ must possess
 norm
smaller than $1/8$ for series to converge.
By adding square roots  $\theta_1=\sqrt{-1},\theta_2= \sqrt{2}$ and
$\theta_3=\sqrt{3}$  and their products one
 obtains 8-dimensional extension. In TGD imbedding space
 $H=M^4_+ \times CP_2$ can be
 regarded locally as 8-dimensional extension of p-adic numbers.
It is probably not an accident that the dimensions of minimal extensions
 allowing square roots are the space time and imbedding space dimensions
 of TGD.

\vm

 By construction any  p-adically real number in the
 extension allows square root.  The square root for an
 arbitrary number sufficiently near real axis can be defined
through Taylor series expansion of the square root function $\sqrt{Z}$ in
point of p-adic
real axis.
 The subsequent considerations show that the p-adic square root function
 does not allow
analytic continuation
 to $R^4$ and the points of extension allowing square root form a set
 consisting of disjoint
converge cubes of square root function forming  structure resembling
lightcone.

\subsection{p-Adic square root function for $p>2$}

 The study of  the properties of the series representation of square
 root function shows that  the definition of square root  function is
possible
in certain region around real p-adic axis. What is nice that this region
 can be
regarded as the  p-adic counterpart of the future light cone
defined by
the condition

\begin{eqnarray}
N_p(Im(Z))&<& N_p(t=Re(Z))=p^k
\end{eqnarray}

\noindent where the real p-adic  coordinate $t=Re(Z)$  is identified as
time
coordinate  and the imaginary    part of the p-adic coordinate is
identified
as spatial coordinate.  p-Adic norm for four-dimensional extension is
analogous  to ordinary Euclidian distance.      p-Adic light cone  consists
of
«cylinders«  parallel  to time axis having radius $N_p(t)= p^k$  and length
$p^{k-1}(p-1)$: at points $t= p^k$.  As a real space (recall the canonical
correspondence)  the cross section of the cylinder corresponds to
parallelpiped
rather than ball.

\vm

The result can be understood heuristically as follows. \\
a) For four-dimensional extension allowing square root  ($p>2$) one can
 construct square root at each p-adically real point $x(k,s)= sp^k$,
$s=1,...,p-1$, $k\in Z$.  The task is to show that  by using Taylor
expansion
one can define square root also in some neighbourhood of each of these
points
and find the form of this neighbourhood.  \\ b)  Using the general series
expansion of the  square root function one finds that the convergence
region is
p-adic ball   defined by the condition

\begin{eqnarray}
N_p(Z-sp^k) &\leq &R(k)
\end{eqnarray}

\noindent and having    radius $R(k) = p^d, d \in Z$ around the expansion
point. \\
 c) A purely p-adic feature is that the   convergence spheres
 associated with
two points are either disjoint or identical!   In particular, the
convergence
sphere $B(y)$ associated with any  point inside convergence sphere
$B(x)$  is
identical with $B(x)$:  $B(y)= B(x)$.  The result follows directly  from
the
ultrametricity  of the p-adic norm.  The result means that
stepwise analytic continuation is not possible and one can construct square
root
function only  in the  union of   p-adic convergence spheres associated
with
the p-adically real points $x(k,s)=sp^k$. \\
d)  By the scaling properties of the square root
function the convergence radius  $R(x(k,s))\equiv R(k)$ is related to
$R(x(0,s))\equiv R(0)$  by the scaling factor $p^{-k}$:

\begin{eqnarray}
R(k) &=& p^{-k}R(0)
\end{eqnarray}

\noindent so that  convergence sphere expands  as a function of p-adic
time
coordinate.   The study of convergence reduces to the study of  the
series at
points $x=s=1,...,k-1$ with unit p-adic norm.  \\ e)  Two neighbouring
points
$x=s$ and $x=s+1$ cannot belong to  same convergence sphere: this would
lead
to contradiction with basic results of about square root function at integer
points.  Therefore the convergence radius  satisfies the condition

\begin{eqnarray}
R(0)&<&1
\end{eqnarray}

\noindent The requirement that  convergence is achieved at all points
 of the
real axis implies

\begin{eqnarray}
R(0)&=&\frac{1}{p}\nonumber\\
R(p^ks)&=& \frac{1}{p^{k+1}}
\end{eqnarray}

\noindent  If the convergence radius is indeed this then the region,  where
square
root is defined corresponds to a  connected light cone like region defined
by the
condition $ N_p(Im(Z))= N_p(Re(Z))$ and
 $p>2$-adic space time is  p-adic counterpart of $M^4$ light cone.  If
 convergence
 radius is smaller the convergence region reduces to a union of disjoint
 p-adic spheres
with increasing radii.

\vm

  How the p-adic light cone differs from the  ordinary light cone
 can
 be seen by studying the explicit form of the p-adic norm
for $p>2$ square root allowing extension $Z=x+iy+\sqrt{p}(u+iv)$

\begin{eqnarray}
N_p(Z) &=& (N_p(det(Z)))^{\frac{1}{4}}\nonumber\\
&=& (N_p((x^2+y^2)^2+2p^2((xv-yu)^2+
(xu-yv)^2)+p^4(u^2+v^2)^2))^{\frac{1}{4}}\nonumber\\
\
\end{eqnarray}

\noindent where $det(Z)$ is the determinant of the linear map defined by
multiplication with $Z$.  The definition of convergence sphere for $x=s$
reduces to

\begin{eqnarray}
N_p(det(Z_3))&=&N_p(y^4+2p^2y^2(u^2+v^2)+p^4(u^2+v^2)^2))<1
\end{eqnarray}

\noindent  For physically interesting case $p \ mod \  4=3$ the points
$(y,u, v)$ satisfying the conditions

\begin{eqnarray}
N_{p}(y)&\leq&\frac{1}{p}\nonumber\\
N_p(u)&\leq& 1 \nonumber\\
N_p(v)&\leq &1
\end{eqnarray}

\noindent belong to the sphere of convergence:  it is essential that
 for all $u$ and $v$ satisfying the conditions one has also
 $N_p(u^2+v^2)\leq
1$.   By the canonical correspondence between p-adic and real numbers
 the real
counterpart of the sphere $r=t$ is now parallelpiped $0\leq y<1,0\leq
u<p,0\leq v<p$,  which expands with average velocity of light in discrete
steps
at times $t=p^k$.

\vm

 The   emergence of  p-adic light cone as a  natural  p-adic coordinate
 space
 is in nice  accordance with the basic assumptions about the imbedding
 space of
TGD and shows that  big bang cosmology might basically related to
the existence
of p-adic
square root!  The result gives
also  support for the idea that p-adicity is responsible for the
generation of lattice structures (convergence region for any function is
expected to be more or less parallelpiped like region).

\vm

A peculiar  feature of the  p-adic light cone is the instantaneous expansion
of 3-space at moments $t_p= p^k$.  A possible physical interpretation is
that  p-adic light cone  or rather the  individual convergence cube of
the light
cone represents the time development of single maximal
quantum coherent region at p-adic level of topological condensate
(probably
there are many of them).  The instantaneous scaling of the size of region by
factor $\sqrt{p}$ at moment
 $t_R= p^{k/2}$ corresponds to a  phase transition and thus  to
 quantum
 jump.    Experience with p-adic QFT indeed shows that $L_p=\sqrt{p}L_0$
,$L_0\simeq 10^4\sqrt{G}$ appears as
 infrared cutoff length for  the p-adic version of standard model so that
p-adic
 continuity is
replaced with real continuity (implying p-adic continuity) above the length
scale $L_p$. The idea
that larger
 p-adic length
scales $p^{k/2}L_p$, $k>0$,   would form quantum coherent regions for
physically most
interesting  values of
$p$ is probably unrealistic and $L_p$ probably gives a typical size of
3-surface
at p:th
condensate level. Of course, already this hypothesis is far from trivial
since $L_p$ can have
arbitrarily large values so that arbitrarily large quantum coherent systems
would be possible.

\vm

   $p=M_{127}$,  the largest physically interesting Mersenne prime,
 provides an interesting example:\\
i) $p=M_{127}$-Adic light cone does not make sense for time and length
scales
smaller than length scale defined by electron Compton length and QFT
below this length
scale makes sense.  \\ ii) The
first phase transition would happen at time of order $10^{-1}$ seconds,
which
corresponds to length scales of order $10^7 $ meters.\\ iii) The next phase
transition takes place at  $t_R\simeq 10^{11}$
light years and corresponds to the age of the Universe.

\vm

Recent work with the p-adic field theory limit of TGD has shown that the
convergence cube of
p-adic
square root function having size $L_p= \frac{L_0}{\sqrt{p}}$,
$L_0= 1.824\cdot 10^4\sqrt{G}$,
serves as a natural quantization volume for p-adic counterpart of
standard model.   An
open question is whether also larger convergence cubes serve as quantization
volumes or
whether $L_p$ gives natural upper bound for the size of p-adic 3-surfaces.
The original
idea  that  p-adic
manifolds, constructed by gluing together pieces of p-adic light cone
together along
their
sides, could be used to build Feynmann graphs with lines thickened to
4-manifolds  has
turned out to be not useful for physically most interesting (large)
values of $p$.

\subsection{Convergence radius for square root function}

In the following it will be shown that the convergence radius of
$\sqrt{t+Z}$ is
indeed nonvanishing for $p>2$.  The expression for the Taylor series of
$\sqrt{t+Z}$ reads as

\begin{eqnarray}
\sqrt{t+Z}&=& = \sqrt{x}\sum_n a_n\nonumber\\
a_n&=& (-1)^n\frac{(2n-3)!!}{2^nn!} x^n\nonumber\\
x&=&\frac{Z}{t}
 \end{eqnarray}

\noindent The necessary criterion for the convergence is that the terms
of the power
series approach to zero at the limit $n\rightarrow \infty$.    The p-adic
norm of
$n$:th term is
for $p>2$ given by

\begin{eqnarray}
N_p(a_n)&=& N_p(\frac{(2n-3)!!}{n!}) N_p(x^n)<N_p(x^n)N_p(\frac{1}{n!})
\end{eqnarray}

\noindent The dangerous term is clearly the $n!$ in the denominator.  In
the following
it will be shown that the condition

\begin{eqnarray}
U&\equiv &\frac{N_p(x^n)}{N_p(n!)} <1 \ for \ \ N_p(x)<1
\end{eqnarray}

\noindent holds true. The strategy is as follows:\\
a) The norm  of $x^n$ can be calculated trivially:
$ N_p(x^n) =p^{-Kn}, K\ge 1$.\\
b)   $N_p(n!) $ is  calculated and an upper bound for $U$ is derived at
the limit of large $n$.

\subsubsection{p-Adic norm of $n!$ for $p>2$}

Lemma 1: Let $n= \sum_{i=0}^{k}n(i)p^i$,  $ 0\le n(i)<p$ be the p-adic
expansion of $n$. Then  $N_p(n!)$  can be expressed in the form

\begin{eqnarray}
N_p(n!)&=& \prod_{i=1}^{k} N(i)^{n(i)}\nonumber\\
N(1)&=&\frac{1}{p}\nonumber\\
N(i+1)&=& N(i)^{p-1}p^{-i}
\end{eqnarray}

\noindent  An explicit expression for $N(i)$ reads as

\begin{eqnarray}
N(i)&=& p^{-\sum_{m=0}^{i}  m(p-1)^{i-m}   }
\end{eqnarray}

\noindent Proof: $n!$ can be  written as a product

\begin{eqnarray}
N_p(n!) &=& \prod_{i=1}^{k}X(i,n(i) )\nonumber\\
X(k,n(k))&=& N_p((n(k)p^k)!)\nonumber\\
X(k-1,n(k-1))&=&N_p(\prod_{i=1}^{n(k-1)p^{k-1}}(n(k)p^k+i))=
N_p( (n(k-1)p^{k-1})!)\nonumber\\
X(k-2,n(k-2))&=&N_p(\prod_{i=1}^{n(k-2)p^{k-2}}(n(k)p^k+n(k-1)p^{k-1}+i
))\nonumber\\
&=& N_p((n(k-2)p^{k-2})!)\nonumber\\
X(k-i,n(k-i))&=& N_p((n(k-i)p^{k-i})!)
\end{eqnarray}

\noindent  The factors $X(k,n(k))$ reduce in turn
to the form

\begin{eqnarray}
X(k,n(k))&=& \prod_{i=1}^{n(k)}Y(i,k )\nonumber\\
Y(i,k)&=& \prod_{m=1}^{p^k} N_p(ip^k+m)
\end{eqnarray}

\noindent The factors $Y(i,k)$ in turn are indentical and one has

\begin{eqnarray}
X(k,n(k))&=& X(k)^{n(k)}\nonumber\\
 X(k)&=& N_p(p^k!)
\end{eqnarray}

  The recursion formula for the factors $X(k)$ can be derived by
writing explicitely the expression of $N_p(p^k!)$ for a few lowest values
of
$k$:\\
  1) $X(1)= N_p(p!) = p^{-1}$\\
2) $ X(2) = N_p(p^2!)= X(1)^{p-1}p^{-2} $ ( $p^2!$ decomposes to  $p-1$
 products having same norm as $p!$ plus the last term equal to $p^2$.\\
i) $ X(i)= X(i-1)^{p-1}p^{-i}$

\vm

Using the recursion  formula repeatedly the explicit form of  $X(i)$ can
be derived easily. Combining the results one obtains for $N_p(n!)$ the
expression

\begin{eqnarray}
N_p(n!)&=& p^{-\sum_{i=0}^{k}n(i) A(i)}\nonumber\\
A(i)&=& \sum_{m=1}^{i} m(p-1)^{i-m}
\end{eqnarray}

\noindent  The sum $A(i)$  appearing in the exponent  as the coefficient of
 $n(i)$ can be calculated by using geometric series

\begin{eqnarray}
A(i)&=& (\frac{p-1}{p-2})^2
(p-1)^{i-1}(1+ \frac{i}{(p-1)^{i+1}}- \frac{(i+1)}{(p-1)^i})\nonumber\\
&\leq& (\frac{p-1}{p-2})^2(p-1)^{i-1}
\end{eqnarray}

\subsubsection{Upper bound for $N_p(\frac{x^n}{n!})$ for $p>2$}

By using the expressions $n= \sum_i n(i)p^i$, $N_p(x^n)=p^{-Kn}$ and the
expression of $N_p{n!}$ as well as the upper bound

\begin{eqnarray}
A(i)
&\leq& (\frac{p-1}{p-2})^2(p-1)^{i-1}
\end{eqnarray}

\noindent for $A(i)$ one obtains the  upper bound

\begin{eqnarray}
N_p(\frac{x^n}{n!}) \leq
p^{-\sum_{i=0}^{k}n(i) p^i(K-(\frac{(p-1)}{(p-2)} )^2
(\frac{(p-1)}{p})^{i-1})} \nonumber\\
 \
\end{eqnarray}

\noindent It is clear that for $N_p(x)<1$  that is $K\ge1$ the upper
 bound goes to zero. For $p>3$ exponents are negative for all values of
 $i$:
for $p=3$ some lowest exponents have wrong sign but this does not spoil the
convergence.  The convergence of the series is also obvious since  the
real valued series $\frac{1}{1-\sqrt{N_p(x)}}$ serves as majorant.

\subsection{$p=2$ case}

In $p=2$ case the norm of a general term in the series of the square root
function can be calculated easily using the previous result for the norm
of $n!$:

\begin{eqnarray}
N_p(a_n)&=& N_p(\frac{(2n-3)!!}{2^nn!}) N_p(x^n)=  2^{-(K-1)n+\sum_{i=1}^{k}
n(i)\frac{i(i+1)}{2^{i+1} }}
\end{eqnarray}

\noindent At the limit $n\rightarrow \infty$  the sum term appearing in
the exponent
approaches zero
and  convergence condition gives $ K>1$  so that one has

\begin{eqnarray}
N_p(Z)&\equiv& (N_p(det(Z)))^{\frac{1}{8}}\leq\frac{1}{4}
\end{eqnarray}

\noindent  The result does  not imply disconnected set of convergence for
 square root function since the square root for half odd integers exists:

\begin{eqnarray}
\sqrt{s+\frac{1}{2}}= \frac{\sqrt{2s+1}}{\sqrt{2}}
\end{eqnarray}

\noindent  so that one can develop square as series in all half odd integer
 points of p-adic real axis.   As a consequence the structure for the set of
convergence is just the 8-dimensional counterpart of the p-adic light cone.
Spacetime has natural binary structure in the sense  that each
$N_p(t)=
2^k$ cylinder consists of two identical p-adic 8-balls (parallelpipeds as
real
spaces).  Since $\sqrt{Z}$ appears in the definition of the fermionic
Ramond
fields  one might wonder whether once could  interpret  this binary
structure as
a geometric representation of  half odd integer spin.  The coordinate
space
associated with spacetime  representable  as a four-dimensional subset
of this
light cone inherits the light cone structure.

\subsection{ p-Adic inner product and Hilbert spaces}

Concerning the physical applications of complex p-adic numbers the problem
is that p-adic norm
is not bilinear in its arguments and therefore it does not define inner
product and angle.
One can however  consider a  generalization of the ordinary  complex inner
product
$\bar{z} z$
to p-adic valued inner product. It turns out that p-adic quantum mechanics
in the
sense as it is
used in p-adic TGD can be based on this inner product.

\vm

  Restrict the consideration to minimal extension
allowing square roots near real axis  ($p>2$) and
 denote the complex conjugate of $Z$ with $Z_c$  and by $\hat{Z}$ the
 conjugate of $Z$ under
 the  conjugation
$\sqrt{p}\rightarrow -\sqrt{p}$: $Z\rightarrow \hat{Z} = x+\theta x-
\sqrt{p}(u+\theta v)$.   The inner
product in the  4-dimensional extension  of p-adic numbers
reads as

\begin{eqnarray}
\langle Z,Z\rangle &=& Z_cZ+ \hat{Z}_c\hat{Z}
=2(x^2+y^2+ p(u^2+v^2))
\end{eqnarray}

\noindent This inner product is bilinear and symmetric, defines p-adically
real norm  and vanishes only if $Z$ vanishes.
This inner product leads
to  p-adic generalization of unitarity and probability concept.
The solution of the unitarity condition
 $\sum_k S_{mk}\bar{S}_{nk}= \delta (m,n)$
involves square root operations and therefore the minimal
extension for the Hilbert space is $4$-dimensional in $p>2$ case
 and $8$-dimensional in $p=2$ case.  The physically most interesting
consequences of this result  are encountered in   p-adic quantum mechanics.

\vm

The inner product associated with minimal extension allowing square root
near real axis
provides a natural  generalization  of the real and complex Hilbert
spaces
respectively.
Instead of real or complex numbers square root allowing algebraic
extension extension
appears as
the multiplier field of the Hilbert space and one can understand the
points of Hilbert
space as
infinite sequences $(Z_1,Z_2,...,Z_n,....) $, where $Z_i$   belongs  to
the extension.
The inner
product $\sum_k \langle Z^1_k,Z_2^k\rangle $ is   completely analogous to
the ordinary Hilbert
space  inner product.

\vm

A particular example of   p-adic Hilbert space is obtained as a
generalization
 of  the space of complex valued functions
$f: R^n\rightarrow C$.  The
  inner product for functions  $f_1$ and $f_2$ is
 just the previously defined inner product
 $\langle f_1(x) ,f_2(x)\rangle$  combined with
integration over $R_p^n$: the definition of p-adic integration will be
considered later
in detail.

\vm

The inner product  allows to define the   concepts of length and
angle for two vectors in p-adic extension possessing
either p-adic or ordinary complex values. This implies that
 the concepts   and of p-adic Riemannian metric, K\"ahler
metric  and conformal invariance  become possible.

\subsection{p-Adic Numbers and Finite Fields}

  Finite fields
(Galois fields) consists of finite number of elements and
 allow sum, multiplication and division.  A convenient
 representation for the elements of a finite field is as
 the roots of the polynomial equation
$t^{p^m}-t=0 \ mod \ p$ , where $p$ is prime, $m$ an
arbitrary  integer and $t$ is element of a field of
 characteristic
$p$ ($pt=0$  for each $t$).   The number of elements in finite field is
$p^m $,
that is power of prime number and  the multiplicative
 group of a finite field is group of order $p^m-1$.
$G(p,1)$ is just cyclic group $Z_p$ with respect to addition and
$G(p,m)$ is in rough sense
$m$:th Cartesian power of $G(p,1)$ .

\vm

  The elements of the finite field $G(p,1)$  can be identified as
the p-adic
numbers  $0,...,p-1$   with p-adic arithmetics replaced with modulo
p arithmetics.
Finite
fields $G(p,m)$ can be  obtained from m-dimensional algebraic extensions
of p-adic
numbers by
replacing p-adic arithmetics with modulo p  arithmetics.  In TGD context
only the
finite fields
$G(p>2,2)$ , $p \ mod \ 4=3$  and $G(p=2,4)$  appear  naturally.
For $p>2$, $p \ mod \ 4 =3$
one has: $x+iy+\sqrt{p}(u+iv) \rightarrow x_0+iy_0 \in G(p,2) $.

\vm

As far as applications are considered the basic observation is that
the     unitary representations of p-adic scalings
 $x\rightarrow p^kx$ $k \in Z$ lead naturally to finite field structures.
These representations  reduce to representations of finite cyclic group
$Z_m$ if
 $x\rightarrow p^m x$ acts trivially on representation functions
 for  some value of $m$,  $m=1,2,..$.  Representation functions,  or
 equivalently the
 scaling momenta  $k=0,1,...,m-1$ labeling them,  have a structure
of
 cyclic group.  If $m \neq p $  is prime the scaling momenta form finite
 field $G(m,1)=Z_m$ with respect to summation and multiplication modulo $m$.
 The construction p-adic field theory shows that
also the p-adic counterparts of ordinary planewaves
carrying p-adic momenta $k=0,1...,p-1$
 can be given the structure of  Finite Field $G(p,1)$: one can also define
complexified planewaves
 as
square roots of the real p-adic planewaves to obtain Finite Field  $G(p,2)$.

\section{p-Adic differential calculus}

It would be nice to have a  generalization of the
ordinary differential and integral calculus  to p-adic
 case.    Instead of
trying to guess  directly the  formal  definition
of p-adic differentiability it is better to guess what kind of
functions $f:R_p \rightarrow R$ might be natural candidates for
p-adically differentiable functions and then try to find whether the
concept of
p-adic differentiability makes sense. There are several candidates for
p-adically
differentiable functions. \\
\noindent a)   p-Analytic maps  $R_p \rightarrow R_p$  representable
as power
series of p-adic argument  induce  via the canonical identification maps
$R_+ \rightarrow R_+$.
These maps are well defined for algebraic extensions of p-adic numbers,
too and induce
p-analytic
maps $R_+^n \rightarrow R_+^n$ via the canonical correspondence. These
functions
correspond to
ordered fractals.\\ b) A second candidate  is obtained as a
generalization of canonical
identification map $R_p \rightarrow R$: $Y_D(x)= \sum x_k p^{-kD}$,
where $D $ is so call
anomalous dimension.   The corresponding  map $R\rightarrow R$ is given
by $\sum_k x_k p^{-k}
\rightarrow \sum  x_k p^{-kD}$: $D=1$ gives identity map. These functions
are not
differentiable in the strict sense of the word and  give rise to
 chaotic fractals, which resemble Brownian functions.

\subsection{p-Analytic maps}

p-analytic maps $g: R_p \rightarrow R_p$ satisfy the usual criterion of
differentiability and
are representable as power series

\begin{eqnarray}
g(x)&=& \sum_k g_k x^k
\end{eqnarray}

\noindent  Also negative powers are in principle allowed. The rules of
p-adic differential
calculus are formally identical to those of the ordinary differential
calculus and generalize
in
trivial manner for algebraic extensions.

\vm

The class of  p-adically constant functions (in the sense that p-adic
derivative vanishes)
is larger than in real case:  any function depending on finite number
of positive pinary
digits of p-adic number and of arbitrary number of negative pinary digits
is p-adically
constant. This becomes obvious, when one considers the definition of
p-adic derivative:
when
the increment of p-adic coordinate becomes sufficiently small p-adic
constant doesn't detect
the
variation of $x$ since it depends on finite number of positive p-adic
pinary digits only.
p-adic
constants correspond to real functions, which are constant below some
length scale $\Delta x=
2^{-n}$.   As a consequence   p-adic differential equations are
nondeterministic:
integration constants are arbitrary functions depending on finite number
of positive p-adic
pinary digits. This  feature  is central as far applications are
considered.

\vm

p-Adically analytic functions induce maps $R_+ \rightarrow R_+$ via the
canonical
identification map. The simplest manner to get some grasp on their
properties is to
plot  graphs
of some simple functions (see Fig. \ref{square} for the graph of p-adic
$x^2$ and for
Fig. \ref{oneoverx}) for the graph of p-adic $1/x$). These  functions
have quite characteristic features resulting from the special properties
of p-adic
topology:\\ a)
p-Analytic functions are  continuous and differentiable from right:
this peculiar
asymmetry    is  a completely  general signature of p-adicity. As far as
time dependence is
considered the interpretation of this property as mathematical counterpart
of irreversibelity
looks  natural. This suggests that the  transition from reversible
microscopic dynamics to
irreversible macroscopic dynamics  corresponds to the transition from the
ordinary topology to
effective p-adic topology.
 \\
b) There are  large  discontinuities associated with the points
$x=p^n$. This implies
characteristic theshold phenomena. Consider a system whose output $f(n)$
is function of input,
which is integer $n$. For $n<p$ nothing peculiar happens but for $n=p$
the real counterpart of
the output becomes very small for large values of $p$. In biosystems
threshold phenomena are
typical and p-adicity might be the key in their understanding. The
discontinuities
associated
with powers of $p=2$ are indeed encountered in many physical situations.
 Auditory experience has the
property that given
frequency $\omega_0$ and its multiples $2^k \omega_0$, octaves,  are
experienced as same
frequency suggesting the auditory response function for a given
frequency $\omega_0$ is
2-adicallly analytic function. Titius-Bode law states that the mutual
distances of planets come in  powers of $2$, when
suitable unit of distance is used. In turbulent systems
 period doubling spectrum  has peaks at frequencies
$\omega = 2^k \omega_0$. \\
c) A second signature of p-adicity
is  "p-plicity"   appearing in the graph of simple p-analytic
functions. As
an example, consider the graph of p-adic $x^2$ demonstrating clearly
the decomposition into $p$ steps at each interval $[p^k,p^{k+1})$.
  \\ d) The graphs of p-analytic functions are in
general  ordered fractals as
 the examples demonstrate. For example, power functions $x^n$ are
 selfsimilar (the
 values of the function at some any interval $(p^k,p^{k+1})$ determines
 the function
completely) and in general p-adic $x^n$ with nonnegative (negative) $n$
is smaller (larger)
  than real $x^n$ expect at points $x= p^n$ as the graphs of p-adic  $x^2$
  and $1/x$ show
 (see Fig. \ref{oneoverx})  These properties are easily understood from the
 properties of
 p-adic multiplication.   Therefore the first guess for the behaviour of
 p-adically analytic
 function is obtained by replacing $x$ and the coefficients $g_k$ with
 their  p-adic norms:
 at points $x= p^n$ this approximation is exact if the coefficients of
 the power series are
 powers of $p$. This step function approximation is rather  reasonable
 for simple
 functions
such as $x^n$ as the figures demonstrate. Since p-adically analytic function
can be approximated  with   $f(x)\sim f(x_0)+b(x-x_0)^n$ or as $a(x-x_0)^n$
(allowing nonanalyticity at $x_0$)  around any point the fractal associated
with p-adically analytic function has universal geometrical form  in
sufficiently small length scales.

\vm

%\begin{figure}
%\leavevmode
%\centering
%\vspace*{1cm}
%\epsfxsize=15 cm \epsfysize=15 cm \epsfbox{square.eps}
%\label{square}
%\caption{p-Adic $x^2$  function for some values of $p$}
%\end{figure}

%\newpage

%\begin{figure}
%\leavevmode
%\centering
%\vspace*{1cm}
%\epsfxsize=15 cm \epsfysize=15 cm \epsfbox{oneoverx.eps}
%\label{oneoverx}
%\caption{p-Adic $1/x$  function for some values of $p$}
%\end{figure}

\vm

 p-Adic analyticity is well defined for the algebraic extensions of
$R_p$, too.
 The figures \ref{real} and \ref{imag} visualize the behaviour of
real and imaginary parts of two adic $z^2$ function as function of real
$x$ and $y$ coordinates in the parallelpiped $I^2$,$I= [1+2^{-7},2-2^{-7}]$.
  An interesting  possibility is that the
order parameters describing various
phases of physical system are p-adically differentiable functions.
  The
p-analyticity would therefore provide a means for coding the information
about
ordered fractal
structures.

\vm

 The order parameter  could be   one
coordinate component
of a
p-adically analytic map  $R^n\rightarrow R^n$, $n=3,4$. This is analogous
to the
possibility to
regard the solution of Laplace equation in 2 dimensions as a real or
imaginary part
of an
analytic function.  A given region  $V$ of the order parameter space
corresponds to
a given phase
and the volume of ordinary space occupied by this phase corresponds to
the inverse image
$g^{-1}(V)$ of $V$. Very beautiful images are obtained  if
the order parameter is the the real or imaginary part of p-adically analytic
function $f(z)$. A good example is
 p-adic $z^2$ function
in the parallelpiped
$[a,b]\times [a,b]$, $a=1+2^{-9}$, $ b=2-2^{9}$ of  $C$-plane.  The value
range of the  order parameter can divided into, say,  $16$ intervals of same
length  so that  each interval corresponds to a unique color.   The resulting
 fractals  possess    features, which probably generalize to
higher dimensional extensions. \\  a) The inverse image is ordered fractal and
possessess lattice/ cell like structure,
with the
sixes of cells appearing in powers of $p$. Cells are however not identical
in analogy
with   the
differentiation of biological cells.    \\ b) p-Analyticity implies  the
existence of
 local
vector valued order parameter  given by the p-analytic derivative of
$g(z)$: the
geometric
structure of the phase portrait  indeed exhibits the local orientation
clearly.  \\
c)  In a
given resolution there appear   0,1, and 2-dimensional structures and
also
defects inside
structures.   In 3-dimensional situation rather rich structures are to
be expected.

\vm

Even more beautiful structures are obtained by adding some disorder:
for instance,  the composite map  $z(x,y)= Y_D(x^2-y^2) $ for $D=1/2$  for
$p=2$, where   the function $Y_D(x)$ is defined in the next section gives
rise to extremely beatiful fractal using the previous description.
 Noncolored pictures  cannot reproduce   the   beauty of these fractals not
suggested  by  the expectations based on the  appearence of the  graphs of
$Y_D(x)$ (see Fig. \ref{yydee}) (the MATLAB
programs needed to generate p-adic  fractals are supplied by request for
interested reader).

\vm

These observations suggests that  p-analyticity might provide a means
to code the
information
about ordered fractal structures  in the spatial behaviour of order
parameters
(such as
entzyme concentrations in biosystems).  An elegant manner to achieve
this is
to use purely
real algebraic extension for 3-space coordinates and for the order
parameter: the
image of the
order parameter $ \Phi= \phi_1+ \phi_2\theta +\phi_3\theta^2$ under the
canonical
identification
is real and positive number automatically and might be regarded as
concentration
type quantity.

%\begin{figure}
%\leavevmode
%\centering
%\vspace*{1cm}
%\epsfxsize=15 cm \epsfysize=15 cm \epsfbox{real.eps}
%\label{real}
%\caption{ The graph  of the real part of 2-adically analytic  $z^2=
%$  function.} \end{figure}
%\newpage

%\begin{figure}
%\leavevmode
%\centering
%\vspace*{1cm}
%\epsfxsize=15 cm \epsfysize=15 cm \epsfbox{imag.eps}
%\label{imag}
%\caption{ The graph of 2-adically analytic  $Im(z^2)= 2xy$  function.}
%\end{figure}

\subsection{Functions $Y_D$}

p-Analytic functions give rise to ordered fractals. One can find also
functions
describing
 disordered fractals.
 The simplest generalization of the
identification of real and p-adic numbers to
a chaotic fractal  is the following one

\begin{eqnarray}
Y_D(x_p)&=&  \sum_n  x(n) p^{-nD}
\end{eqnarray}

\noindent where $D$ is constant.  $D=1$ gives identication map.  $Y_D$
defines
in obvious
manner a map $R_+\rightarrow R_+$ via the canonical identification map.
To each pinary digit
there corresponds the power $p^{-kD}$ so that a change of single
pit induces  change of form $p^{-kD}$  nonlinear in the increment
$\vert dx_p \vert $.
\vm

One can generalize the definition of $Y_D$ . The
anomalous dimension $D$ can be p-adic constant and therefore  depend on
finite number
of positive  p-
adic pinary digits of $x_p$ and the most general definition of $Y_D$
reads

\begin{eqnarray}
Y_D(x_p)&=&  \sum_n  x(n) p^{-nD(x_{<n+1})}\nonumber\\
D&=& D(x_p)\nonumber\\
x_{<n} &=& \sum_{k<n} x_kp^k
\end{eqnarray}

\noindent Here it is essential to assume that  the anomalous
dimension
associated with n:th pinary digit is the value of anomalous dimension
associated with n:th pinary digit cutoff of $x$: otherwise p-adic
continuity
is lost. This generalization allows also fractal  functions, which
become ordinary smooth functions in sufficiently small length scales:
the only assumption needed is that $D(x)$ approaches $D=1$, when the
number of pinary digits of $x$ becomes large.  The definition of the
functions $Y_D$ generalizes in trivial manner to higher dimensional case,
the anomalous
dimensions being now p-adically constant functions of all p-adic
coordinates.

\vm

Although the functions $Y_D$ are  not differentiable in the strict sense of
the word
 they have the property that if $x_p$ has
finite
number of nonvanishing  pinary digits then for sufficiently small
increment
$dx_p$ so that $x_p$ and $dx_p$ have no common pinary digits  one has
just

\begin{eqnarray}
Y_D(x_p+dx_p)-  Y_D(x_p) &=& Y_D (dx_p)
\end{eqnarray}

\noindent  $Y_D(dx_p)$ might be called $D$-differential with
anomalous dimension $D$. This differential  maps the tangent
space of p-adic numbers to the tangent space of real numbers in
fractal like  manner in the sense that if p-adic and real tangent
spaces are identified in canonical manner then $D$-differential
induces nonlinear fractal like map of real tangent space to itself.
The functions $Y_D(dx_p)$ are therefore good candidate for a fractal
like generalization
of
linear differential $dx_p$.

\vm

The local anomalous dimension $D$ corresponds to the
so called
 Lifschitz-H\"older  exponent  $\alpha$ encountered in the
theory of
multifractals \cite{Frac}. Multifractals are decomposed into union of
fractals with various fractal dimensions by decomposing the range
$S$ of fractal function to a union $ \cup_{\alpha} S_{\alpha}$ of
disjoint sets $S_{\alpha}$ : $S_{\alpha}$ consists of points of $S$,
for which the anomalous dimension $D$ has fixed value $\alpha$. One
can associate to each set $S_{\alpha}$ its own fractal dimension and
this decomposition plays important role in fractal analysis of the
empirical data. The values of $D>1$ and $D<1$ one correspond to
"antifractal" (ordinary derivative vanishes) and fractal (ordinary
derivative is divergent) behavior respectively.
\vl

The simplest manner to see the  fractality properties is
to plot the graph of  $Y_D$.  The general features of the graph
(see. Fig. \ref{yydee} )
are following:\\
a) $Y_D$ is continuous from right and there are  sharp
discontinities
associated
with the points $x= p^m$. The graph of $Y_D$ is  selfsimilar
if $D$ is
constant. The
value of $p$ reflects itself as a characteristic  "p-peakedness"
for $D<1$
and
"p-stepness" for $D<1$. \\ b)  For $D<1$ $Y_D$ is surjective but not
injective.
This is
seen as
typical fluctuating behaviour resembling that associated with Brownian
motion.
If $D<1$ is
constant there is infinite number of preimages associated with a given
point $y$. \\
c) For $D>1$
$Y_D$ is  constant almost everywhere, nonsurjective,  and increases
monotonically.
\\
d) $x=0$ and $x=1$ are fixed points common to all  $Y_D$. These
points are attractors for $D<1$ and repellors for $D>1$.
If $D<1$ $Y_D$ has also additional fixed points $x>1$ in the
neighbourhood $x=1$.

\vl

%\begin{figure}
%\leavevmode
%\centering
%\vspace*{1cm}
%\epsfxsize=15 cm \epsfysize=15 cm \epsfbox{yydee.eps}
%\label{yydee}
%\caption{ The graph  of the   function $Y_D(x)$ for various values of
%$D$ and $p$.}
%\end{figure}

The graph of $Y_D$, $D<1$ resembles that of  Brownian motion. The
following
arguments
suggest that there is more than a mere analogy involved and that functions
$Y_D$ with  p-adically
constant $D$ combined with ordinary differentiable functions might provide
a description
for
random processess. \\ a) $Y_D$ equals to $p^{kD}$ at points $x=p^D$. For
$D=1/2$ this
means that
$Y_D$ is analogous to the root mean square distance
$d(t)=\sqrt{\langle r^2\rangle } (t)$
from
origin in Brownian motion, which behaves as $ d \propto \sqrt{t}$.  \\
b) In Brownian
motion
$d(t)$ is not differentiable function at origin: $d(t) \propto t^{1/2}$.
The same holds
true for
$Y_D$,$D=1/2$ at each point $x$ so that $Y_D$ in certain sense provides a
simulation of
Brownian
motion.   \\ c) $Y_D$ is only the simplest example of Brownian looking
motion and as
such too
simple to describe realistic situations.  It is  however possible to form
composites of
$Y_D$ and
p-adically differentiable functions as well as ordinary differentiable
functions, which
both are
right differentiable with anomalous dimension $D=1$.  These functions
contain also p-adic
constants, which depend on finite number of pinary digits of $t$ in
arbitrary manner so
that
nondeterminism results. These features suggest that  the functions
$Y_D$ provide
basic
element for the description of Brownian processess. \\ d) There is no
obvious reason to
exclude
values of $D$ different from $D=1/2$ and this means that the concept of
Brownian motion
generalizes. \\ e) Since Brownian motion can be regarded as Gaussian
process (the value
of the
increment of $x$ obeys Gaussian distribution) it seems that also higher
dimensional
Gaussian
processes possessing as their graphs Brownian surfaces could be described
by using the
higher
dimensional algebraic extensions of p-adic numbers and corresponding higher
dimensional
extensions of $Y_D$. The deviation of $D$ from $D=1/2$ might correspond to
anomalous
dimensions
deriving from the non-Gaussian behaviour implied by interactions.

\vm

One can form also functional composites of $Y_D$ and anomalous dimensions
are multiplicative
in this process: $y_{D_1} \circ y_{D_2}$ possessess anomalous dimension
$D= D_1 \times D_2$.
For $D_i <1$ the functional composition in general implies more chaotic
behaviour.  It must
be emphasized, that functions $Y_D$  (not very nice objects!) do not have
appear in any
applications of this book.

\section{ p-Adic integration}

The concept of p-adic  definite integral can be defined for functions
 $R_p\rightarrow  C$ \cite{padrev} using translationally invariant  Haar
 measure
for $R_p$.  In present context one is however interested in definining
p-adic
valued  definite integral for functions $f: R_p\rightarrow R_p$: target
and
source spaces could of course be  also some some algebraic extensions of
p-adic
numbers.
  What makes the definition nontrivial  is that the ordinary
definition as the limit of Riemann sum doesn«t work:  Riemann sum
approaches to
zero in p-adic topology and one must somehow circumvent this difficulty.
Second
difficulty
is related to the absence of well ordering for p-adic numbers.  The
problems are
avoided by
defining integration essentially as the inverse of differentation and using
canonical
correspondence to define ordering for p-adic numbers.

\vm

The definition of p-adic integral functions defining integration as inverse
of
differentation operation
  is straightforward and one obtains just the generalization of standard
calculus. For instance,  one has $\int z^n = \frac{z^{n+1}}{(n+1)}+ C$ and
integral of Taylor series is obtained by generalizing this.   One
must however notice that the concept of  integration constant generalizes:
  any
function $R_p\rightarrow   R_p$ depending on finite number of pinary digits
only, has vanishing derivative.

\vm

Consider next definite integral.    The absence of well
ordering  implies that the concept of integration
 range $(a,b)$ is not well defined as purely p-adic concept.   A possible
resolution of the problem is based on canonical identification.  Consider
p-adic
numbers $a$ and $b$. It is natural to define $a$ to be smaller than $b$ if
the
canonical images of $a$ and $b$ satisfy $a_R<b_R$.   One must notice that
$a_R=b_R$ does not imply $a=b$ since the inverse of the canonical
identification
map is two-valued for real numbers having finite number of pinary digits.
For
two p-adic numbers $a,b$ with  $a<b$ one can define the integration range
$(a,b)$ as the set of p-adic numbers $x$ satisfying $a\leq x\leq b$ or
equivalently $a_R\leq x_R\leq b_R$.   For a given value of $x_R$  with
finite
number of pinary digits one has two values of $x$ and $x$ can be made unique
by requiring it to have finite number of pinary digits.

\vm

   One can  define
definite integral $\int_a^b f(x)dx$  formally as

\begin{eqnarray}
\int_a^b f(x)dx&=& F(b)-F(a)\nonumber\\
\end{eqnarray}

\noindent  where $F(x)$ is
integral function obtained by allowing only ordinary integration constants
and
$b_R>a_R$ holds true.   One encounters
however problem,  when $a_R=b_R$  and $a$ and $b$ are different. Problem
is
avoided if integration limits are assumed to correspond p-adic  numbers with
finite number of pinary digits.

\vm

One could perhaps relate the possibility of p-adic integration constants
depending on
finite
number of pinary digits to the possibility to decompose integration range
$[a_R,b_R]$ as
$a=x_0<x_1<....x_n=b$ and to select in each subrange $[x_k,x_{k+1}]$ the
inverse images
of $x_k\leq x\leq x_{k+1}$, with  $x$ having finite number of pinary digits
in two
different
manners. These different choices correspond to different integration paths
and the value
of
the integral for  different paths could correspond to the different choices
of p-adic
integration constant in integral function. The difference between a given
integration
path and
'standard' path is simply the sum of differences $F(x_k)-F(y_k)$,
$(x_k)_R= (y_k)_R$.

\vm

This definition has several nice features: \\
a) Definition generalizes in obvious manner to higher dimensional case. \\
g) Standard connection between integral function and definite integral
holds
true and in higher dimensional case the integral of total divergence
reduces to
integral over boundaries of integration volume. This property guarantees
that
p-adic action  principle leads to same field equations as its real
counterpart. It
this in fact
this property, which drops other alternatives from consideration. \\
c) Integral is linear operation and additive as a set function. \\
d) The basic  results of real integral calculus generalize as such to p-adic
case.

\vm

There is however a problem related to the generalization of the integral
to the case of non-analytic functions.  For instance,  the so called number
theoretic plane waves defined as functions $a^{kx}$ with $a\in \{1,p-1\}$ is
so called primitive root satisfying $a^{p-1}=1$ and   $k \in
Z$,   are p-adic counterparts of ordinary plane waves and nonanalytic
functions
of $x$.  The construction of field theory limit of TGD is based on Fourier
analysis
using p-adic planewaves. It is difficult to avoid the use of these
functions in
construction of p-adic version of  perturbative QFT.  What one needs is
definition of
integral guaranteing orthogonality of the p-adic  plane waves
in suitable
integration range. A formal integration using the integration formula
gives factor

\begin{eqnarray}
\int_0^{p-1} a^{kx}dx&=&  \frac{1}{ln(a)}(a^{k(p-1)}-1)=0, \ k=1,...,p-1,
\nonumber\\ \int_0^{p-1}a^{kx}dx_{\vert k=0} &=& \int_0^{p-1} dx = p-1
\end{eqnarray}

\noindent Although the factor $ln(a)$ is ill defined p-adically this does
not
matter since integral vanishes for $k\neq 0$: for $k=O$ the integral is in
well defined.
p-Adic planewaves are not differentiable in ordinary sense but the
differentation  can be
defined purely algebraically as multiplication with p-adic momentum.

\section{ p-Adic manifold  geometry}

In the following the concepts of  p-Adic Riemannian  and conformal
geometries are
considered.

\subsection{p-Adic Riemannian geometry}

   It is possible to  generalize the concept of (sub)manifold  geometry to
   p-adic
   (sub)manifold geometry.
The formal definition of p-adic Riemannian geometry is based on p-adic line
element
$ds^2= g_{kl} dx^kdx^l$. Lengths and angles are defined in  the usual
manner and their
definition involves  square root $ds$  of the line element. The
existence of
square roots forces quadratic extension of p-adic numbers  allowing square
roots. As found
the extension is $4$-dimensional for $p>2$ and $8$-dimensional in $p=2$
case.   This
extension in question must appear as coefficient ring of p-adic tangent
space so that
  p-adic Riemann spaces  must  be  locally cartesian powers  of $4-$
($p>2$) or
8-dimensional ($p=2$)  extension. Therefore  spacetime and imbedding
space
dimensions of TGD emerge very naturally in p-adic context.

\vm

The definition of pseudo-Riemannian metric poses problem:  it seems that
one
 should be able to make distinction between negative and
 positive p-adic numbers.  A possible manner to make this distinction is
 to p-adic numbers with unit norm to be positive or
negative according to whether they are squares or not.  This definition
 makes sense if $-1$ does not possess square root: this is true for
 $p \ mod \ 4 = 3$.  This condition will be  encountered in most
 applications of p-adic
 numbers.
 At analytic level the definition generalizes in obvious manner: what is
 required that
 the components of the metric are p-adically real numbers. The p-adic
 counter part of the
Minkowski metric can be  defined as

\begin{eqnarray}
ds^2_p  &=& (dm^0)^2 - ((dm^1 )^2+(dm^2 )^2+(dm^3 )^2)
\end{eqnarray}

\noindent The real image of this line element under canonical
identification is
nonnegative but metric  allows to define the p-adic counterpart
of $ M^4$ lightcone as the
surface $(m^0)^2 - ((m^1 )^2+(m^2 )^2+(m^3 )^2)=0$ and this surface can
be regarded as a
fractal counterpart of the ordinary light cone. Furthermore, this
metric allows the
p-adic counterpart of Lorentz group as its group of symmetries.

\vm

An interesting possibility is that one could define the length of a
fractal curve («coast
line of Britain«)  using p-adic Riemannian geometry.  A possible model
of this curve is
obtained by identifying ordinary real plane with its p-adic counterpart
via canonical
identification and modelling the fractal curve with p-adically continuous
or even analytic
curve $x=x(t)$. The real counterpart of this curve is certainly fractal
and need not have
well defined length. The p-adic length of this curve can be defined as
p-adic integral of
$s_p= \int ds$  and its real counterpart $s_R$ obtained by canonical
identification can be
defined to be the real length of the curve.

\vm

The concept of p-adic Riemann manifold as such is not quite enough for the
mathematization  of  the
topological condensate concept. Rather,  topological condensate can be
regarded as a
surface obtained by glueing together p-adic spacetime regions with
different values of $p$
together along their boundaries. Each region is regarded as submanifold
p-adic counterpart
of $H= M^4_+\times CP_2$. A natural manner to perform the gluing operation
is to use
canonical identification to map the  boundaries of two regions $p_1$
and $p_2$ to  real
imbedding space $H$ and  to require that $p_1$ and $p_2$ boundary points
correspond to
same point in $H$.

\subsection{ p-Adic conformal geometry}

It would be nice to have a generalization of ordinary conformal
 geometry to p-adic context.   The following considerations and results of
 p-adic TGD
 suggest that the  induced K\"ahler form defining Maxwell field on
 spacetime surface
could  be the basic entity of 4-dimensional conformal geometry rather
than metric.   If
the existence of square root is
 required the dimension of this geometry is $D=4$ of $D=8$ depending
 on the value of $p$.  In the following it is assumed that the
 extension used is the minimal extension allowing square root and
 $p \ mod \ 4=3$
 condition holds so that imaginary unit belongs to the generators of
 the extension.

\vm

In 2-dimensional case  line element   transforms by a conformal scale
factor in p-analytic map $Z\rightarrow f(Z)$.  In four-dimensional case
this requirement
leads to degenerate line element

\begin{eqnarray}
ds^2&=& g(Z,Z_c,...) dZdZ_c \nonumber\\
&=& g(Z,Z_c,..) (dx^2+dy^2+p(du^2+dv^2)+ 2\sqrt{p}(dxdu+dydv))
\end{eqnarray}

\noindent where the conformal factor $g(Z,Z_c,..)$ is invariant under
 complex conjugation.   The metric tensor associated with the line element
 does not
 possess inverse.
 This is obvious from the fact that line element depends on two coordinates
$Z,Z_c$ only so that p-adic conformal metric is effectively 2-dimensional
rather than 4-dimensional.   It therefore seems that one must give up
conformal
covariance requirement for  line element.

\vm

In two-dimensional  conformal geometry angles are simplest conformal
invariants
 and are expressible in terms of  the inner product.   In 4-dimensional
 case
 one can  define invariants, which are analogous to angles.
 Let $A$ and $B$ be two vectors in 4-dimensional  quadratic extension
 allowing square root.  Denote A (B) and its various conjugates
 by $A_i$ ($B_i$), $i=1,2,3,4$.  Define phase like quantities $X_{ij}=$
 «$exp(i2\Phi_{ij})$«
 between $A$ and $B$ by the following formulas

\begin{eqnarray}
X_{ij}&\equiv &  \frac{A_i A_j B_kB_l }{\sqrt{A_1A_2A_3A_4}
\sqrt{B_1B_2B_3B_4}}
\end{eqnarray}

\noindent where $i,j,k,l$ is permutation of $1,2,3,4$.
Each quantity $X_{ij}$ is invariant under one of the conjugations
${}_c$, $\hat{}$ or $\hat{}_c$ and $X_{ij}$ has values in 2-dimensional
subspace
 of the  4-dimensional extension.
As in ordinary case the angles are invariant under conjugation and this
 means that only $3$ angle like quantities exists: this is in accordance
 with the fact that 3-angles are needed to specify the orientation of the
 vector $A$ with respect to the vector $B$.

\vm

One can define also more general invariants using four vectors $A,B,C,D$
and
 permutations $i,j,k,l$ and $r,s,t,u$ of $1,2,3,4$

\begin{eqnarray}
U_{ijkl}&=& \frac{X_{ijkl}}{X_{rstu}}\nonumber\\
X_{ijkl}&\equiv&A_iB_jC_kD_l
\end{eqnarray}

\noindent   The number of the functionally independent invariants is
reduced if various
conjugates of invariants are not counted as different invariants.
If 2 or 3 vectors are
identical one obtains as special case invariants associated with 3 and
2 vectors.  If
there are only two vectors  the number
 of the functionally independent invariants is  $6$.

\vm

There exists  quadratic conformal covariants associated with tensors of
 weight two.  The general form of the covariant is given by

\begin{eqnarray}
X&=&g^{ij:kl} A_{ij} B_{kl}
\end{eqnarray}

\noindent The tensor $g^{ij:kl}$ has the property that in  complex
coordinates
 $Z,\bar{Z},\hat{Z},\bar{\hat{Z}}$ the only nonvanishing components
 of the
tensor have $i\neq j\neq k\neq l$. This guarantees multiplicative
transformation
property in conformal transformations $Z\rightarrow W(Z)$:

\begin{eqnarray}
X (W) &= & \frac{dW}{dZ}
\frac{d\bar{W}}{d\bar{Z}}
\frac{d\hat{W}}{d\hat{Z}}
\frac{d\bar{\hat{W}}}{d\bar{\hat{Z}}} X(Z)
\end{eqnarray}

\noindent  The simplest example of tensor $g^{ij:kl}$ is permutation symbol
and the instanton density of any gauge field defines p-adic conformal
covariant (the
quantity is  actually $Diff^4$ invariant).

\vm

The  K\"ahler form of $CP_2$ is
self dual but this property in general does not hold true for the
induced K\"ahler form
defining Maxwell field on spacetime surface.
  K\"ahler action density  (Maxwell action) formed formed
from the induced K\"ahler form on spacetime surface  is in general not
p-adic
conformal invariant as such whereas the 'instanton density'  is conformal
invariant.
It turns out that if $CP_2$  complex  coordinate (4-dimensional extension)
is
p-adically analytic  function of $M^4$ complex coordinate then  the induced
K\"ahler form
is self dual in the approximation that the induced metric is flat and one
can express
K\"ahler action density as

\begin{eqnarray}
J^{\alpha\beta}J_{\alpha\beta}&=& \epsilon^{\alpha\beta\gamma\delta}
J_{\alpha\beta}J_{\gamma\delta}
\end{eqnarray}

\noindent This quantity satisfies the conditions guaranteing multiplicative
transformation property under p-adic conformal transformations. What is
nice that
p-dically analytic maps define approximate
extremals of K\"ahler action:  action density however vanishes identically
in flat metric
approximation. An interesting open problem is whether one could find
more general extremals
of K\"ahler action  satisfying the condition

 \begin{eqnarray}
 J^{\alpha\beta}&=& g^{\alpha\beta\gamma\delta}J_{\gamma\delta}
 \end{eqnarray}

 \noindent such that the tensor $g^{...}$  satisfies the required
 conditions but does not
reduce to the permutation symbol.

\vm

 Whether 4-dimensional  p-adic conformal invariance plays  role in p-adic
 TGD is not
clear.  It turns out the entire $Diff^(M^4)$ rather than only $Conf(M^4)$
acts as
approximate symmetries of K\"ahler action (broken only by gravitational
effects) and that
it is this larger invariance, which seems to be  relevant for the dynamics
of the interior
of spacetime surface. It is p-adic counterpart  of the ordinary
2-dimensional conformal
invariance  on boundary components of 3-surface, which plays key role
in the calculation
of particle  masses.

\section{p-Adic symmetries}

The most basic level questions physicist can ask about
 p-adic numbers are related to symmetries.
It  seems  obvious that the  concept of
Lie-group generalizes: nothing prevents
from replacing the real or complex representation spaces
associated with the definitions of classical Lie-groups
 with linear space associated with some algebraic extension of p-adic
 numbers: the
 defining
 algebraic conditions,
 such as unitarity  or orthogonality properties, make
 sense for algebraically extended p-adic numbers, too.
In case of orthogonal groups one must replace the ordinary real inner
product with
 p-adically real inner product $\sum_k X_k^2$ in  a Cartesian power
of a purely real
extension of p-adic numbers: it should be emphasized that this inner
product must be
p-adic
valued.  In the unitary case one must consider complexification of a
Cartesian power of
purely
real extension  with inner produc $\sum \bar{Z_k} Z_k$.   It should be
emphasized however
that
the  p-adic inner product differs from the ordinary one so that the
 action of, say,
p-adic
counterpart of rotation group in $R_p^3$ induces in $R^3$ an action, which
need not have
much to
do with ordinary rotations  so that the generalization is physically
highly nontrivial.
For
very large values of $p$ there are however good reasons to expect that
locally the action
of
these groups resembles the action of their real counter parts.

\vm

A simple example is provided by the generalization of rotation group
$SO(2)$. The rows of
a rotation matrix are in general $n$ orthonormalized vectors with the
property that the
components of these vectors have p-adic norm not larger than one. In
case of $SO(2)$ this means
the the matrix elements $a_{11}=a_{22}=a,a_{12}=-a_{21}=b$ satisfy the
conditions

\begin{eqnarray}
  a^2+b^2&=&1 \nonumber\\
\vert a\vert_p &\leq &1 \nonumber\\
\vert b\vert_p &\leq &1
\end{eqnarray}

\nonumber One can formally solve $a$ as $a=\sqrt{1-b^2}$ but the solution
doesn't exists always.
There are various possibilities to define the orthogonal group.\\
a) One possibility is to allow only those values of $a$ for which square
root exists as p-adic
number.  In case of orthogonal group this requires that both
$b= sin(\Phi)$ and
$a=cos(\Phi)$
exist as p-adic numbers.  If one requires rurther that  $a$ and $b$ make
sense also  as
ordinary
rational numbers, they define  Pythagorean triangle with integer sides
and the group
becomes
discrete and cannot be regarded as Lie-group.   Pythagorean triangles
emerge for any
rational
counterpart of  any Lie-group.   \\ b)  Other possibility is to allow an
extension of
p-adic
numbers allowing square root. The minimal extensions has dimension $4$ (8)
for $p>2$
($p=2$).
Therefore spacetime dimension  and imbedding space  dimension emerge
naturally as
minimal
dimensions for
 spaces,  where p-adic $SO(2)$ acts 'stably'.  The requirement that $a$
 and $b$ are
 real  is necessary unless one wants complexification of the $so(2)$ and
 gives
 constraints on
the values of group parameters and again Lie-group property is expected to
be lost. \\
 c) Lie-group property is guaranteed if the allowed group elements are
 expressible as
 exponents of Lie-algebra generator $Q$. $g(t)= exp(iQt)$. This exponents
 exists only
provided the p-adic norm of $t$ is smaller than one.  If one uses square
root allowing
extension
one can require that $t$ satisfies $\vert t\vert \leq p^{-n/2}$, $n>0$ and
one obtains a
decreasing  hierarchy of groups $G_1,G_2,..$.   For physically interesting
values of $p$
(typically of order $p=2^{127}-1$ )  the real counterparts of the
transformations of
these groups
are extremely near to unit element of group.  These conclusions hold true
for
any group. An especially interesting example physically is the group of
'small'  Lorentz
transformations  with $t=O(\sqrt{p})$. If the rest energy
of the particle is of order   $O(\sqrt{p})$: $E_0= m= m_0\sqrt{p}$ (as it
turns out) then the
Lorentz boost with velocity $\beta= \beta_0\sqrt{p}$  gives particle with
energy $E=
m/\sqrt{1-\beta_0^2p}= m(1+\frac{\beta_0^2p}{2}+..)$ so that $O(p^{1/2})$
term in energy is
Lorentz invariant. This suggests that   nonrelativistic regime corresponds
to  small
Lorentz transformations whereas in genuinely relativistic regime  one must
include also
the discrete group of 'large'  Lorentz transformation with rational
transformations matrices.

{}

\end{document}